# Room temperature detection of the $(H_2)_2$ dimer

H. Fleurbaey, S. Kassi, A. Campargue

*Univ. Grenoble Alpes, CNRS, LIPhy, Grenoble, France*

Corresponding author
E-mail address: Alain.Campargue@univ-grenoble-alpes.fr
Tel.: 33 4 76 51 43 19   Fax. 33 4 76 63 54 95

**Abstract**

The hydrogen dimer, $(H_2)_2$, is among the most weakly bound van der Waals complexes and a prototype species for first principles *ab initio* studies. The detection of the $(H_2)_2$ infrared absorption spectrum was reported more than thirty years ago at a temperature of 20 K. Due to the sharp decrease of the $(H_2)_2$ abundance with temperature, a detection at room temperature was generally considered as hardly achievable. Here we report the first room temperature detection of partly resolved rotational structures of $(H_2)_2$ by cavity ring down spectroscopy at sub-atmospheric pressures, in the region of the first overtone band of $H_2$ near 1.2 µm. The quantitative analysis of the absorption features observed around ten allowed or forbidden transition frequencies of the monomer provides insight on the structure of this elusive species and a benchmark for future theoretical studies.

**1. Introduction**

In the spectroscopic databases commonly used for atmospheric or planetary applications (*e.g.* the HITRAN database [1]), the hydrogen absorption spectrum is obtained as the sum of two contributions: the very weak electric quadrupole transitions of the $H_2$ molecule showing up as distant narrow rovibrational lines and the collision induced absorption (CIA) bands with a very broad and smooth structure mostly coinciding with the range of the (*v*-0) $H_2$ overtone bands (*v* is the vibrational quantum number). Let us recall that the absorption of the $H_2$ lines is proportional to pressure while the CIA pressure dependence is quadratic. The possible additional contribution due to the $(H_2)_2$ dimer is assumed to be negligible at atmospheric temperatures and thus neglected.

With a binding energy of about 3 cm$^{-1}$ (~ 4.5 K), the hydrogen dimer, $(H_2)_2$ is indeed among the most weakly bound van der Waals complexes and thus expected to have an extremely low concentration at room temperature. In order to theoretically predict its absorption spectrum, the multi-dimensional potential energy surface (PES) of the $(H_2)_2$ dimer must be calculated with an accuracy significantlybetter than 1 cm$^{-1}$, which represents a real challenge for first principles quantum chemical calculations. Started in the 1940s and 1950s [2], [3] [4], the calculations of the interaction potential between two $H_2$ molecules remains nowadays an active field [5], [6], [7], [8], [9], [10], [11], [12], [13]. Beyond its academic interest, the $H_2$ dimer is of special interest in various domains. As hydrogen is more and more considered as a versatile energy carrier with important needs for storage and transportation at standard temperature conditions, the equilibrium thermodynamics of the $H_2$ dimers in high pressure tanks of hydrogen is particularly relevant in the present period. In planetary science, the $(H_2)_2$ spectral signature has been evidenced in the far-infrared (FIR) spectra of Jupiter and Saturn recorded in the Voyager IRIS mission [14], [15], [16], [18]. The assignment to $(H_2)_2$ of the faint narrow structure observed in planetary spectra near the S(0) and S(1) pure rotational lines was confirmed by McKellar from laboratory absorption spectra of $H_2$ at 77 K [15], [17], [18].

Although the geometry of the hydrogen dimers has been very recently visualized using Coulomb explosion imaging [19], absorption spectroscopy appears to be a complementary method to provide observations and validation tests for theoretical calculations. Due to the weakness of the binding energy, all previous spectroscopic detection of the $H_2$ dimer were reported at low temperature. The $H_2$ dimer contribution to the absorption spectrum of hydrogen at 20 K was first demonstrated in 1964 by Watanabe and Welsh in the region of the (1-0) fundamental band of $H_2$ centered near 4150 cm$^{-1}$ [20]. Later, McKellar and Welsh recorded the same band at a similarly low temperature with a grating spectrograph and an improved spectral resolution and partly assigned the few weak sharp dimeric features surrounding the electric quadrupole $H_2$ lines [21]. A further improvement of the quality of the $(H_2)_2$ infrared spectrum in terms of spectral resolution and sensitivity was achieved in [22] by Fourier transform spectroscopy (FTS) with a spectral resolution in the range of 0.04-0.16 cm$^{-1}$. The FTS recordings were performed at 20 K and pressure values of a few tens Torr and extended up to the first overtone region, (2-0), for $H_2$, $D_2$ and HD, and their mixtures. In summary, the review of the literature indicates that the previous spectra of the $H_2$ dimer were recorded at low temperature and are limited to rotational transitions at 77 K in the FIR and spectra at 20 K in the fundamental band. To the best of our knowledge, no experimental spectra of the $H_2$ dimer were reported since 1991.

Revisiting the $(H_2)_2$ spectrum with modern highly sensitive laser-based techniques appears to be timely. The sensitivity nowadays achieved by cavity-enhanced techniques allows for the quantitative detection of trace gases with extremely low concentration (see *e.g.* [23]). In the present work, we adopted the cavity ring down spectroscopy (CRDS) technique and investigated the high sensitivity room temperature absorption spectrum of hydrogen in the 8000-8500 cm$^{-1}$ region of the (2-0) first overtone band. The investigated region includes the $Q_2$(1-3) and $S_2$(0) electric quadrupole lines. In his study of the (1-0) band at 20 K (and 25 Torr) [22], McKellar observed a few sharp lines assigned to transitions between bound states of the dimer around the $H_2$ quadrupole lines. Our aim is to investigate to which extent similar sharp lines are detectable and preserved at room temperature in the (2-0) band, taking into account the expected sharp decrease of the dimer concentration and the possible change of the appearance of the dimeric rotational structure.

**2. Experimental set-up**

The CIA and the electric quadrupole lines of $H_2$ in the 8000-8500 cm$^{-1}$ spectral region under study are presented in **Fig. 1**. Only four electric quadrupole lines of the (2-0) first overtone band are located in the region. Following Watanabe&Welsh [20] and McKellar [22], we label them $Q_2$(1), $Q_2$(2), $Q_2$(3), and $S_2$(0) (the number 2 in subscript referring to the $v$= 2 value of the upper vibrational state). The corresponding line centers and line intensities are given in **Table 1**.

The present experimental study follows two previous contributions by cavity ring down spectroscopy (CRDS) dedicated to the $H_2$ absorption spectrum in the region. In Ref. [24], the frequency of the electric quadrupole transition lines was determined with sub-MHz accuracy. The zero-pressure transition frequencies were obtained by applying a multi-spectrum fit procedure with various beyond-Voigt profile models to a series of spectra recorded at different pressures over a spectral interval limited to less than 0.5 cm$^{-1}$ around the line center. As illustrated below, this spectral amplitude is insufficient to reveal the dimer spectral signature which extends over several cm$^{-1}$ around the center of the $H_2$ electric quadrupole lines. A second contribution was devoted to CIA measurements performed during pressure ramps up to 1 atm, at 28 spectral points selected far from absorption lines [25]. A purely quadratic pressure dependence was obtained for the CRDS absorption coefficient at each measurement point and the CIA binary coefficients derived with a 1.5 % accuracy were found significantly higher than the HITRAN values (see **Fig. 1**) (Note that HITRAN CIA values are calculated values from [26]).

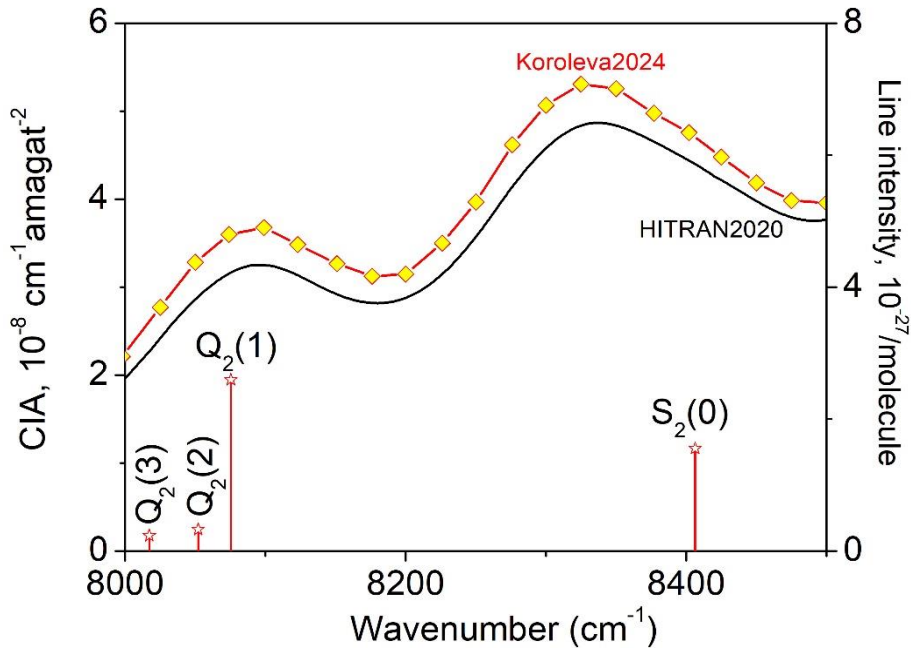

**Fig. 1.**
The quadrupole lines and CIA in the 8000-8500 cm$^{-1}$ region of the first overtone band of H$_2$. The binary coefficients (left-hand scale) as recommended by the HITRAN database (black line) or measured by CRDS [Koroleva2024] (yellow diamonds) are superimposed to the stick spectrum of the (2-0) electric quadrupolar transitions (right-hand scale).

The CRDS setup used for the present recordings is similar to that used in [25]. Briefly, an external cavity diode laser (ECDL) (Toptica fiber-connected DL pro, 1200 nm) tunable over the region of interest (8000-8650 cm$^{-1}$) is used as light source and injected in a 1.4-meter-long high-finesse cavity. For the evacuated cavity, the ring-down time $\tau_0$ varied from about 160 μs to 340 μs depending on wavenumber. For each frequency step, the average central emission frequency of the ECDL was actively stabilized using a software based Proportional-Integral loop, with 100 Hz bandpass, acting on the laser current. For each frequency point, about 40 ring-downs were averaged leading to a minimum detectable absorption coefficient between $2\times10^{-12}$ and $8\times10^{-12}$ cm$^{-1}$ for a single scan. Following the frequency comb referenced CRDS method (CR-CRDS), each ring-down event was associated to its laser frequency. A self-referenced frequency comb (Model FC 1500-250 WG from Menlo Systems) was used for the frequency calibration of the spectra. The laser frequency is measured "on the fly" by recording the beat note between a fraction of the ECDL light and the nearest tooth of the frequency comb, the tooth number being determined using a Fizeau type wavemeter (HighFinesse WSU7-IR, 5 MHz resolution, 20 MHz accuracy over 10 hours) (see [27] and [28] for details).

The temperature measured with an analog temperature sensor (TSic 501, IST-AG, ±0.1 K accuracy) fixed on the cell surface varied between 292.6 and 294.5 K according to the recordings.

The CR-CRDS absorption spectra of natural dihydrogen (Alphagaz2, 99.9999 % chemical purity) were recorded in flow regime in order to reduce spectral interferences with very strong lines of water vapor desorbing from the CRDS cell or from the injection tubes. The hydrogen pressure in the CRDS cell was actively regulated to set values through a needle valve connecting the cell to a turbo pump group and a voltage-controlled valve connecting the cell to the hydrogen cylinder, using a computer based Proportional/Integral controller. The gas pressure in the cell was continuously monitored by a capacitance gauge (MKS Baratron, 1000 mbar full range).

The spectra were recorded for series of pressure values ranging between 10 and 1000 Torr over about 10 cm$^{-1}$ around the $Q_2(1)$, $Q_2(2)$, $Q_2(3)$, and $S_2(0)$ electric quadrupole lines - see **Table 1**. Additional recordings were performed around the calculated frequencies of the $Q_2(0)$ transition, which is forbidden for symmetry reasons, and several collision-induced transitions involving a double excitation of two H$_2$ molecules either both in the (1-0) fundamental band: $Q_1(1) + Q_1(0-3)$ or the pure rotational transition $S_0(0)$ combined with overtone transitions: $S_0(0) + Q_2(0-1)$. The corresponding frequency values obtained from [29] are listed in **Table 1**.

**Table 1.**
Summary of the spectral intervals of the CR-CRDS recordings in the 8000-8500 cm$^{-1}$ region of the (2-0) band of H$_2$.

|  | Position (cm$^{-1}$) [29] | Intensity (cm/molecule) [29] | Pressure (Torr) | | | | | | |
| --- | --- | --- | --- | --- | --- | --- | --- | --- | --- |
|  |  |  | 50 | 100 | 200 | 350 | 500 | 750 | 1000 |
| $Q_2(3)$ | 8017.183 | 2.36×10$^{-28}$ |  |  |  |  |  | × |  |
| $Q_2(2)$ | 8051.988 | 3.29×10$^{-28}$ | × | × | × | × |  | × |  |
| $Q_2(1)$ | 8075.307 | 2.60×10$^{-27}$ | × | × | × | × | × | × | × |
| $Q_2(0)$ | 8087.003 | - | × | × | × | × |  | × | × |
| $Q_1(1)+Q_1(3)$ | 8281.126 | - |  |  | × | × |  | × |  |
| $Q_1(1)+Q_1(2)$ | 8298.719 | - |  | × | × | × |  | × |  |
| $Q_1(1)+Q_1(1)$ | 8310.507 | - |  | × | × | × |  | × |  |
| $Q_1(1)+Q_1(0)$ | 8316.420 | - |  | × | × | × |  | × |  |
| $S_2(0)$ | 8406.361 | 1.56×10$^{-27}$ |  |  |  |  | × | × |  |
| $Q_2(1)+S_0(0)$ | 8429.681 | - |  |  |  |  | × | × |  |
| $Q_2(0)+S_0(0)$ | 8441.376 | - |  |  |  |  |  | × |  |

### 3. Detection of the (H$_2$)$_2$ spectrum

As illustrated in **Fig. 2**, even in flow regime, the residual concentration of water vapor (estimated to be between 10 and 30 ppm in most cases, up to 500 ppm for lower H$_2$ flow rate) is sufficient to have the spectra dominated by water rovibrational lines (broadened by H$_2$). This is due to the very strong transitions of the second hexad of H$_2$O located in the region which have an intensity up to 5×10$^{-23}$ cm/molecule i.e. four orders of magnitude larger than the strongest (2-0) H$_2$ lines, the dimer absorption being itself orders of magnitude weaker than the H$_2$ lines. The procedure adopted to get rid of the water interference lines uses the HITRAN water vapor line list as starting point. The water vapor spectrum was fitted using a Voigt profile and subtracted from the CRDS spectrum. In the process, the water concentration was adjusted together with the line center and the broadening coefficient which are both affected by the H$_2$ buffer gas. The result of the subtraction of the water vapor contribution over the 8000-8420 cm$^{-1}$ region is presented on the lower panel of **Fig. 2** for a pressure recording of 760 Torr. The resulting H$_2$ absorption is dominated by the CIA which shows two maxima in the region and by the $Q_2(1)$, $Q_2(2)$, $Q_2(3)$, and $S_2(0)$ electric quadrupole lines. In addition, at the scale of the plot, unexpected weak absorption features are clearly detected at the bottom of the $Q_2(1)$ and $S_2(0)$ lines and near twice the frequency of the $Q_1(1)$ transition (8310.507 cm$^{-1}$) which is free of electric quadrupole line.

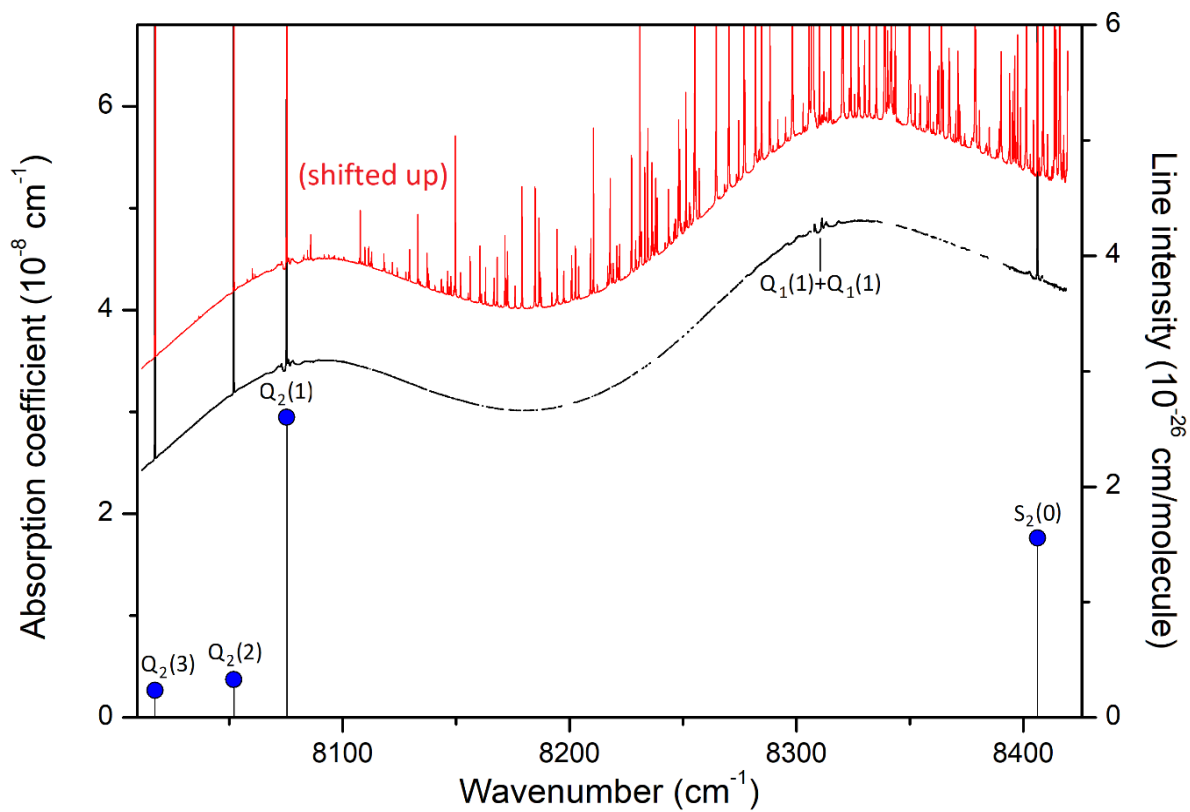

**Fig. 2.**
Overview of the hydrogen spectrum at room temperature ($P=$ 760 Torr) in the 8000-8450 cm$^{-1}$ interval of the first overtone band. The zero-absorption baseline due to the mirror reflectivity has been subtracted. The black curve is obtained after subtraction of the absorption lines of water vapor, visible on the red curve which has been shifted up by 10$^{-8}$ cm$^{-1}$ for clarity. The mole fraction of water was about 25 ppm for this recording. The blue dots show the positions and intensities of the $Q_2(1-3)$ and $S_2(0)$ monomer lines. The calculated position of the $Q_1(1)+Q_1(1)$ transitions which is not radiatively allowed, is also indicated.

A detailed view of the spectrum around the $Q_2(1)$ line before and after removal of the water absorption is presented in **Fig. 3** (upper panel). The absorption features surrounding the H$_2$ line are unambiguously assigned to H$_2$ dimer by comparison to the dimer features observed by McKellar in his FTS spectra in the region of the $Q_1(1)$ line of the (1-0) fundamental band [22]. At first sight, the high similarity between the dimeric structures observed at 20 K and 300 K (and in distinct vibrational bands) is surprising. It will be discussed below.

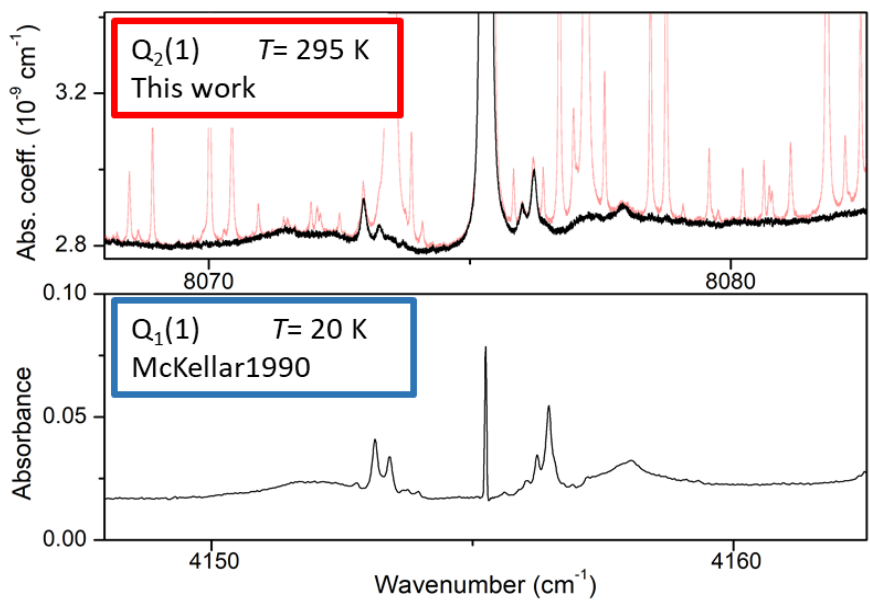

**Fig. 3.**

Comparison of the $(H_2)_2$ absorption spectrum near the $Q_2(1)$ line of the first overtone of $H_2$ and near the $Q_1(1)$ line of the (1-0) fundamental band (lower panel). The frequency axes of the two spectra are centered on the $H_2$ transition frequencies and the width of the displayed spectral interval is identical.

*Upper panel*: Room temperature CR-CRDS spectrum near the $Q_2(1)$ line recorded at 200 Torr. The dimer spectrum (black) is obtained after subtraction of the water vapor contribution (light red),

*Lower panel*: FTS spectrum at 20 K recorded by McKellar in the region of the $Q_1(1)$ line. The $H_2$ pressure was 25 Torr and the absorption pathlength was 112 m [22].

The overview of the different dimeric spectral features observed in the entire region is presented in **Fig. 4.** Spectra recorded at 350 and 750 Torr are superimposed. Overall, ten absorption structures are assigned to the dimer. Four of them are surrounding an electric quadrupole line of the (2-0) band: $Q_2(1-3)$ and $S_2(0)$. The remaining six are located near the sum of the frequencies of two $H_2$ transitions either both in the (1-0) fundamental band, $Q_1(1) + Q_1(0-3)$, or the $S_0(0)$ pure rotational excitation combined with an overtone transition, $S_0(0) + Q_2(0)$, and $S_0(0) + Q_2(1)$. The calculated frequencies of the $H_2$ allowed and forbidden transitions (given in **Table 1**) correspond to the blue vertical lines displayed in **Fig. 4**. It is worth noting that both the relative spectral scale and the intensity scale are the same on the different panels. A roughly estimated baseline due to the CIA was subtracted from each spectrum. Around the four $H_2$ quadrupole lines, the intensity of the dimeric structures appears to roughly scale with the intensity of the line.

For comparison purposes, we give the orders of magnitude of the various absorptions involved. For instance, in the $Q_2(1)$ region at 1 atm, the highest absorption coefficient of the dimer structure is of the order of $10^{-9}$ cm$^{-1}$ to be compared with a peak absorption coefficient of about $10^{-6}$ cm$^{-1}$ for the $Q_2(1)$ line (see Fig. 2 of [30]) and a value of about $3.5 \times 10^{-8}$ cm$^{-1}$ for the CIA absorption level (see **Fig. 1**). The noise equivalent absorption of the recorded CR-CRDS spectra varies from 2 to $8 \times 10^{-12}$ cm$^{-1}$, depending on the spectral region.

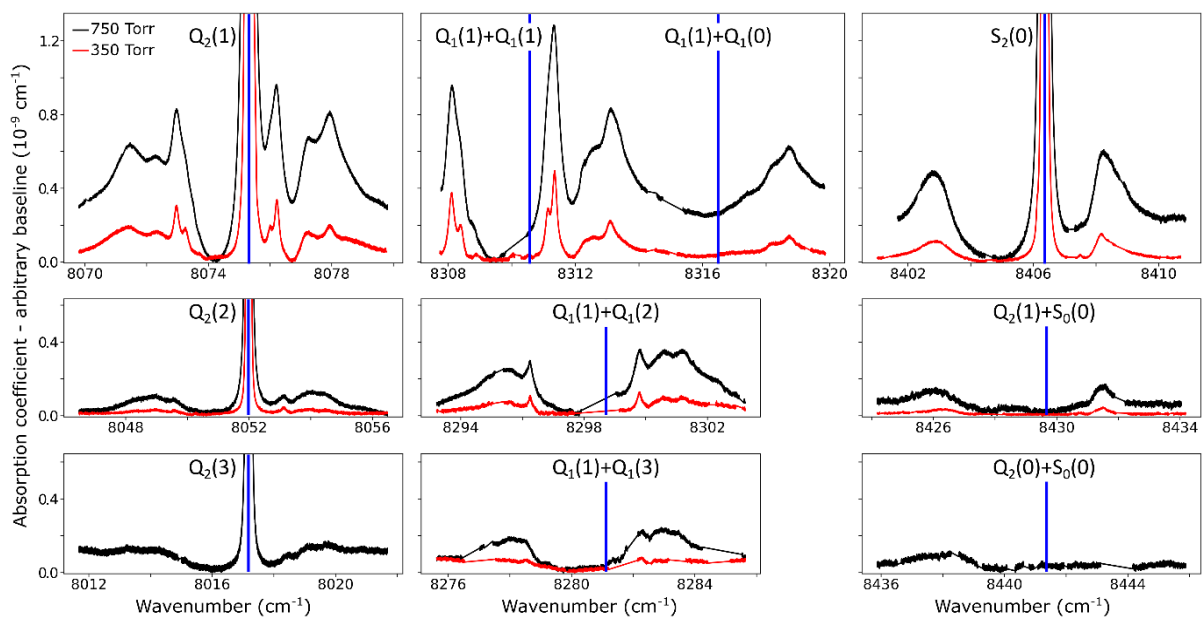

**Fig. 4.**
Close-up view of dimer structures observed around $H_2$ monomer lines and simultaneous transitions, at 350 Torr (red) and 750 Torr (black). Some spectral gaps are observed due to strong absorption lines of water vapor. Calculated positions of monomer resonances are indicated by blue vertical lines. The same relative spectral scale and the same intensity scale are used for the different panels. An empirical linear baseline was subtracted from each spectrum.

The pressure dependence of the $(H_2)_2$ absorption structure is displayed in **Fig. 5** for the $Q_2(1)$ region for which the largest series of recordings were performed (see **Table 1**). A clear broadening of the sharp dimer lines is observed (see also **Fig. 4**).

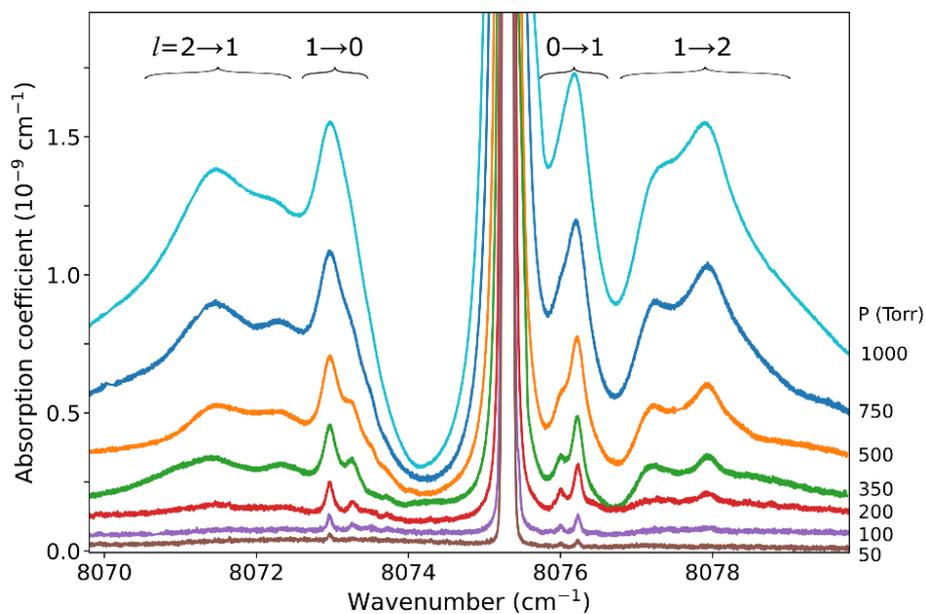

**Fig. 5.**
Pressure dependence of the dimer structure around the $Q_2(1)$ line. An empirical linear baseline was subtracted from the spectra. A partial assignment is indicated above, where $l$ is the end-over-end rotational quantum number of the dimer.

## 4. Quantitative analysis of the $(H_2)_2$ spectrum

At first sight, the similarity of the dimer spectra at 20 K and at room temperature is unexpected (**Fig. 3**). It is due to the fact that only the very first rotational levels in the ground vibrational level ($n=0$) are bound and give rise to sharp lines ($n$ is the vibrational quantum number of the van der Waals stretch). This applies to all temperatures. More precisely, the observed dimer structures arise from the $l=0$ and $l=1$ levels which are bound and from the $l=2$ and $l=3$ levels which are quasi-bound, *i.e.* orbiting resonances bound only by the centrifugal distortion energy barrier [22] ($l$ is the end-over-end rotational angular momentum). Higher rotational levels give rise to very broad absorption features which contribute to the large absorption background and are indistinguishable from the CIA [31].

Only a partial assignment is available for the different absorption features (see **Fig. 5**). Following [7], [22], the different sharp structures observed around $Q_2(1)$, $Q_1(1) + Q_1(1)$, and $Q_1(1) + Q_1(2)$ are assigned to transitions involving the $l=0$ and $l=1$ bound levels, thus 1→0 and 0→1 on the low and high energy side of the monomer transition, respectively. The observation of resolved doublets for both 1→0 and 0→1, near the $Q_2(1)$ and $Q_1(1) + Q_1(1)$ lines, has been interpreted by McKellar as due to the splitting of the total angular momentum, $J$, which is the vector sum of the $l$ end-over-end rotational angular momentum of the dimer with $j$ rotational momenta of the two monomers. This splitting reflects the angular anisotropy of the intermolecular potential [22]. (Note that, as in the fundamental region, a few additional sharp lines significantly weaker than the main doublets are observed – see **Figs. 3** and **6**). Such narrow features are not observed around the $S_2(0)$ line; in the fundamental band, they are predicted to be much weaker due to the different type of dipole inducing mechanisms [21], [32]. Transitions involving higher $l$ quasi-bound states give rise to much broader resonances.

Although the dimer absorption structures appear as a complex superposition of features with distinct shapes and amplitudes, it is worth trying to model them empirically to account for the observations and provide quantitative information for future analysis and comparison to theoretical results. The sharp $l=0 \leftrightarrow l=1$ lines could be modeled as a sum of absorption lines with a Voigt line-shape, the Gaussian component being fixed the Doppler broadening of $(H_2)_2$ at room temperature. An example of such a reproduction of the sharp lines surrounding the $Q_1(1) + Q_1(1)$ simultaneous transition is displayed in **Fig. 6**. The parameters of observed sharp absorption features are listed in **Table 2**. They include the absolute line position and the position value relative to the monomer transition frequency, the Lorentzian width and the normalized integrated absorption coefficient obtained from the ratio of the line area ($A$) by the pressure squared:

The broader, $l=1 \leftrightarrow l=2$ structures, however, have distinctively asymmetric line shapes resembling Fano line. This is particularly striking in the structure surrounding the $S_2(0)$ line (**Fig. 4**). A more detailed analysis of these features is beyond the scope of this paper.

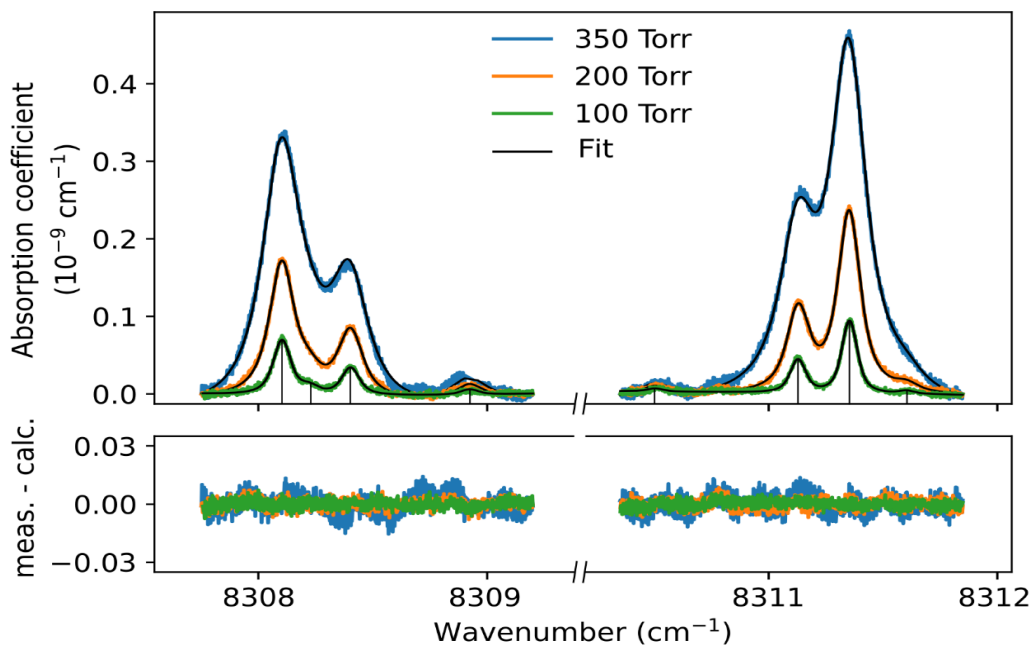

**Fig. 6.** Sharp dimer features around the $Q_1(1) + Q_1(1)$ simultaneous transition. For three pressure values, the observed features (here shown with a linear baseline subtracted) are fitted with a sum of Voigt functions with the Gaussian component fixed to the calculated Doppler broadening value of the dimer.

**Table 2.**
Fitted line shape parameters of the sharp $H_2$ dimer lines.

| Monomer transition | Position (cm$^{-1}$) | Relative position [a] (cm$^{-1}$) | Broadening coefficient (cm$^{-1}$/atm) | $A/P^2$ [b] ($10^{-10}$ cm$^{-2}$/atm²) |
|---|---|---|---|---|
| Q1 | 8072.966 | -2.342 | 0.19 | 2.84 |
| Q1 | 8073.266 | -2.042 | 0.19 | 1.14 |
| Q1 | 8073.716 | -1.592 | 0.19 | 0.36 |
| Q1 | 8075.649 | 0.341 | 0.21 | 0.28 |
| Q1 | 8076.007 | 0.699 | 0.21 | 1.95 |
| Q1 | 8076.229 | 0.921 | 0.21 | 4.48 |
| Q1 | 8076.429 | 1.121 | 0.21 | 0.56 |
| Q1+Q1 | 8308.100 | -2.407 | 0.19 | 4.88 |
| Q1+Q1 | 8308.220 | -2.287 | 0.19 | 0.61 |
| Q1+Q1 | 8308.401 | -2.107 | 0.19 | 2.44 |
| Q1+Q1 | 8308.950 | -1.557 | 0.19 | 0.49 |
| Q1+Q1 | 8310.501 | -0.006 | 0.19 | 0.25 |
| Q1+Q1 | 8311.129 | 0.622 | 0.19 | 2.76 |
| Q1+Q1 | 8311.351 | 0.844 | 0.19 | 6.07 |
| Q1+Q1 | 8311.601 | 1.094 | 0.19 | 0.25 |
| Q1+Q2 | 8296.25 | -2.47 | [c] | [c] |
| Q1+Q2 | 8299.792 | 1.073 | 0.20 | 1.37 |

Notes.
[a] Difference between the measured position and the nearest transition frequency of the monomer (see **Table 1**),
[b] Normalized integrated absorption coefficient determined from the quadratic pressure fit of the integrated absorption coefficients in **Fig. 7** and equal to ratio of the line area by the pressure squared.
[c] This dimer line is partly obscured by a strong water line which hindered the retrieval of reliable line parameters.

The pressure dependence of the integrated absorption coefficient and of the Lorentzian width is shown for five of the sharp lines in **Fig. 7**. Let us note that the observed Lorentzian broadening of all the sharp lines (left panel) is similar and proportional to pressure with no significant broadening at zero-pressure. This confirms the absence of predissociation effects for the $l=0$ and 1 states involved in the considered transitions. Note that the obtained pressure broadening coefficient of $(H_2)_2$ by $H_2$ [about $\gamma_D = 0.20(1)$ cm$^{-1}$/atm (HWHM)] is significantly larger than the self-broadening coefficient of the $H_2$ monomer (about 0.05 cm$^{-1}$/atm [1]). It is interesting to compare $\gamma_D$ to $\gamma_c = 1/(2\pi\tau_c)$ where $\tau_c$ is the average time between two collisions. If each collision of the dimer with an $H_2$ molecule leads to the dissociation of the dimer $\gamma_c$ is expected to be close to $\gamma_D$. The average time between two collisions is the ratio of the mean free path to the average molecular velocity: $\tau_c = \frac{1}{\pi d^2 \overline{v_{rel}} n}$ where $d = r_1 + r_2$ is the effective kinetic diameter ($r_1$ and $r_2$ are the kinetic radii of $(H_2)_2$ and $H_2$, respectively) and $\overline{v_{rel}} = \sqrt{\frac{8kT}{\pi\mu}}$ where μ is the reduced mass of $(H_2)_2$ and $H_2$ and $n$ is the volume density of $H_2$ molecules. The kinetic radius of $H_2$ is 1.445 Å. If we assume that the kinetic radius of the dimer coincides with the Lennard-Jones radius of about 3.4 Å [19], we obtain a broadening coefficient of 0.21 cm$^{-1}$/atm close to the measured value and thus confirming that each collision dissociates the fragile dimer structure.

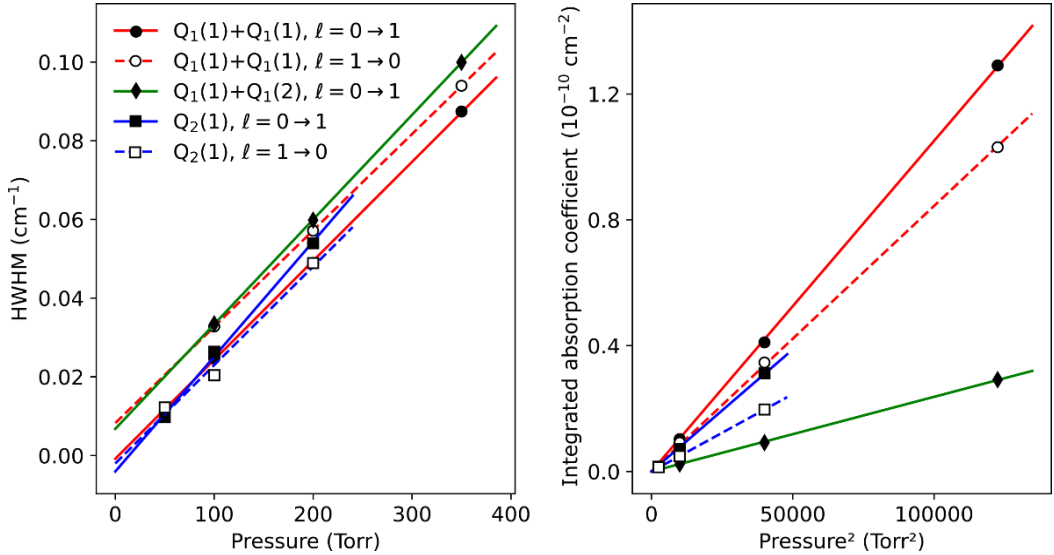

**Fig. 7.** Lorentzian width and integrated absorption coefficient as a function of pressure, from Voigt fits of the largest sharp dimer lines around the $Q_1(1) + Q_1(1)$, $Q_1(1) + Q_1(2)$ and $Q_2(1)$ transitions. The $Q_1(1) + Q_1(2)$, $l=1\rightarrow0$ line was not included because it is obscured by a strong water line. For the resolved doublets, only the largest of the two components is shown here. The other component is weaker by a factor close to 2 and has an identical width within fit uncertainties.

A clear negative position shift is observed for the apparent center of the dimer structure compared to the position of the monomer line. The average position value of the four sharp lines near $Q_2(1)$ and $Q_1(1) + Q_1(1)$ is shifted by -0.6906 and -0.7624 cm$^{-1}$ compared to the corresponding monomer line center, respectively. This value can be compared to a value of -0.44 cm$^{-1}$ obtained for the $Q_1(1)$ transition of the fundamental band [22] - see **Fig. 3**. The negative value of the shift indicates that the dimer potential well is deeper when the monomer is vibrationally excited.

As concerns the separation of the average value of the 1→0 and 0→1 doublets, we obtain a value of 3.00 cm$^{-1}$, which would roughly correspond to 4$B$ in a rigid molecule ($B$ is the dimer rotational constant). From this $B$ value, we derive an average distance between the two H$_2$ molecules of about 4.5 Å which is consistent with the distribution of H$_2$-H$_2$ intermolecular distances recently measured by Coulomb explosion imaging [19]. Interestingly, from the (H$_2$)$_2$ spectra in the fundamental band at 20 K which led Watanabe&Welsh [20] to identify bound states for the first time, sixty years ago, a 4.2-4.6 Å range of intermolecular distances was already proposed.

The measured value of the integrated absorption coefficient is the product of the dimer concentration (in cm$^{-3}$) by the line intensity (in cm/molecule). As the latter is unknown, our quantitative measurements do not allow for an evaluation of the dimer concentration. Recently, starting from the equation of state (EOS), Halpern evaluated the $K_D$(atm$^{-1}$)= $P_{dimer}/(P_{H2})^2$ equilibrium constant of the H$_2$ dimerization, 2H$_2$ ↔ (H$_2$)$_2$, in the 25-45 K temperature range [13]. Statistical thermodynamics (ST) calculations were found consistent with the EOS results. Under our request, Halpern extended his ST calculations up to 300 K and obtained a value of the $K_D$ constant of 3.7×10$^{-6}$ atm$^{-1}$ indicating that, at 1 atm, the relative concentration of the H$_2$ dimer is 3.7 ppm. Note that the room temperature $K_D$ constant is about four orders of magnitude smaller than its 20 K value. As an energy vector, hydrogen gas is currently pressurized in tanks at room temperature and pressures of several hundred atmospheres. Based on the above $K_D$ constant, the dimer relative concentration in these tanks reaches values higher than 0.1%.

**5. Conclusion**

The room temperature detections of dimers or weakly bound complexes are scarce. This is due to their small relative concentration and the difficulty to evidence their weak spectral signature which is frequently obscured by the much stronger absorption of the monomer molecules. The room temperature detection of resolved absorption lines of the water dimer was only possible in the millimeter-wave region where the spectral interferences with the H$_2$O monomer lines are limited [33], [34]. The water dimer contribution to the water absorption continuum in the infrared region has been evidenced based on the correspondence between the observations and calculated band contours of the dimer (see *e.g.* [35]) but the detection and identification of sharp resolved lines are the most convincing way to prove the presence of weakly concentrated complexes in gas mixtures.

Regarding spectral congestion, with its very sparse absorption spectrum, the hydrogen molecule is a more favorable case than water vapor for the detection of its dimer but the very small binding energy of about 3 cm$^{-1}$ (to be compared to a value of 1105 cm$^{-1}$ [36] in the case of the water dimer) leads to much lower relative abundances in the case of (H$_2$)$_2$ (the room temperature equilibrium constant is about four orders of magnitude larger in the case of water dimer [37]).

The comparison of the contribution of the $(H_2)_2$ and $(H_2O)_2$ spectrum to the hydrogen and water vapor absorption, respectively, is interesting. The water absorption is modeled as a sum of two contributions: the rovibrational lines and the absorption (self)-continuum. The semi-empirical MT_CKD model [38], [39] which is the standard model of the water continuum implemented in climate and weather prediction models is a far-wing line shape model of the monomer. The basis of the MT_CKD model is thus not related to water dimer even though it is now established that an important part of the water continuum is due to the water dimer. In the case of molecular hydrogen, the absorption includes also both absorption lines and a continuum (see **Fig. 1**) but the continuum has a different physical origin as the CIA is due to transient electric dipole moments induced by collisions between $H_2$ molecules. The hydrogen CIA values recommended by the HITRAN database are calculated values obtained by neglecting the dimer contribution [26]. At room temperature, broad absorption features of $(H_2)_2$ may extend over several tens cm$^{-1}$ around the $H_2$ transition frequencies and thus contribute to the $H_2$ continuum. The present measurements confirm that the dimer contribution is significantly weaker than the CIA but the most intense sharp lines of $(H_2)_2$ may represent up to 3% of the CIA level (Note that both have a quadratic pressure dependence).

In spite of its weakness, it is necessary to examine to which extend the $(H_2)_2$ spectrum may bias metrological measurements of the $H_2$ quadrupole lines. In recent years, highly accurate measurements of the $H_2$ (and HD) transition frequencies have been used to challenge the most accurate theoretical calculations, including relativistic and quantum electrodynamics (QED) effects (*e.g.* [40]). Experimentally, transition frequencies were determined with a few tens kHz error bars for a few transitions recorded either in saturation regime (Lamb dips) (*e.g.* [41]) or in the Doppler-limited regime (*e.g.* [24]). While Lamb dip measurements are unaffected by dimers, a possible bias on the $H_2$ line centers determined in Doppler regime has to be considered. Although a detailed error budget remains to be performed, from the present observations, it appears that the spectral interference between the $(H_2)_2$ and $H_2$ absorption is too limited to significantly bias the derived $H_2$ transition frequencies.

More than 30 years after the last report on the absorption spectrum of the $(H_2)_2$ dimer and while all previous studies were performed at low temperature (*e.g.* 20 K in the fundamental band), the high sensitivity of the CRDS technique has allowed us to measure the first $(H_2)_2$ spectrum at room temperature (and sub-atmospheric pressures). As only a few low-*l* states give rise to clear absorption features, the observed $(H_2)_2$ spectrum appears to be mostly independent on the temperature.

Empirical parameters were retrieved for the dimeric features observed around ten allowed and forbidden $H_2$ transition frequencies in the region of the first overtone band, near 8000 cm$^{-1}$. We hope that the obtained absolute position and absorption values together with the provided widths of the sharp and broad features will stimulate theoretical efforts and serve as a benchmark for future validation tests. Indeed, in spite of its apparent simplicity, to the best of our knowledge, no first-principles calculations of the $(H_2)_2$ spectrum are available for a direct comparison with our infrared measurements. Due to the very small binding energy of the hydrogen van der Waals dimer and the absence of a preferred geometry, a highly accurate six-dimensional potential energy surface is required. The evaluation of the strengths and dissociation rates of the observed (H$_2$)$_2$ transitions is an additional challenge to predict the spectrum. The first quantitative intensity information on the (H$_2$)$_2$ absorption obtained in this work will be valuable to check the consistency between the $2H_2 \leftrightarrow (H_2)_2$ dimerization constant and the line strengths provided by theory. In the present context of an increasing use of molecular hydrogen as an energy vector, the knowledge of the (H$_2$)$_2$ abundance in highly pressurized tanks used for transportation at room temperature might have some industrial impact.

As a final conclusion, the present work opens new perspectives for direct room temperature detection of dimers or complexes of the major molecular species, with possible impact in the study of the Earth and planetary atmospheres.

*Conflicts of interest*

There are no conflicts to declare.

**Acknowledgements**
We are indebted for useful exchanges and helpful discussions with Robert McKellar who kindly supplied us with his FTS spectrum displayed in Fig. 2, and with several theoreticians: Andrey Vigasin (Moscow), Magnus Gustafsson (Luleå, Sweden) and Arthur M. Halpern (Indiana, USA) who provided us unpublished values of the dimer equilibrium constant which are given in the Discussion. The support by the REFIMEVE consortium (Equipex REFIMEVE+ ANR-11-EQPX-0039) is acknowledged.